\PassOptionsToPackage{table,xcdraw}{xcolor}
\documentclass[10pt,conference]{IEEEtran}

\usepackage[table]{xcolor}  
\usepackage{amssymb}
\usepackage{pifont}

\ifCLASSOPTIONcompsoc
  \usepackage[nocompress]{cite}
\else
  \usepackage{cite}
\fi
\ifCLASSINFOpdf
\else
\fi
\usepackage{amsmath}
\usepackage{balance}  % to better equalize the last page
\usepackage{graphics} % for EPS, load graphicx instead
\usepackage{txfonts}
\usepackage{times}    % comment if you want LaTeX's default font
\usepackage{color}
\usepackage{textcomp}
\usepackage{booktabs}
\usepackage{ccicons}

\usepackage{cite}
\usepackage{url}
\usepackage{fancybox}
\usepackage{multirow}
\usepackage{flushend}
\usepackage{booktabs}
\usepackage{tabularx}
\usepackage{comment}
\usepackage{array}
\usepackage[flushleft]{threeparttable}
\usepackage{mdframed}
\graphicspath{{}{images/}{dia/}}
\DeclareGraphicsExtensions{.pdf,.png}

\usepackage{listings}
\usepackage{courier}
\usepackage{hyperref}

\usepackage{xpatch}

\xpatchcmd{\refstepcounter}{%
  \stepcounter{#1}%
}{%
  \stepcounter{#1}%
  \xdef\scr{\number\value{#1}}%
}{\typeout{success}}{\typeout{failure}}

\newcounter{o}
\usepackage{tikz}
\usepackage{styles/pgf-pie}
\usetikzlibrary{positioning,shadows}
\usepackage{balance}

\newif\ifpienumberinlegend
\pgfkeys{/number in legend/.code=
    \expandafter\let\expandafter\ifpienumberinlegend
    \csname if#1\endcsname
    \ifpienumberinlegend

    \def\beforenumber##1\afternumber{}%
    \fi,
    /number in legend/.default=true
}
\definecolor{1c1}{RGB}{188,162,6}
\definecolor{1c2}{RGB}{137,129,80}
\definecolor{1c3}{RGB}{239,167,31}
\definecolor{1c4}{RGB}{88,194,241}
\definecolor{1c5}{RGB}{6,180,188}

\tikzset{mynode/.style={draw=white,solid,circle,fill=green,inner sep=1pt, thick,
text=black}}
\tikzset{arrow line/.style={dashed, line width= 2.5pt, color=#1}}

\def\bf{\textbf}

\def\it{\textit}

\def\tt{\mct}

\newcommand{\mct}[1]{{\footnotesize {\texttt {#1}}}}

\usepackage{paralist}

\usepackage{caption}
\usepackage{subcaption}

\newcommand{\nd}{\vspace{1mm}\noindent}

\usepackage{tikz}

\makeatletter

\newcommand\notsotiny{\@setfontsize\notsotiny{6}{6}} 

\makeatother

\lstdefinestyle{inlinecode}{basicstyle={\ttfamily\scriptsize\bfseries}}

\newcommand{\urls}[1]{{\scriptsize\url{#1}}}
\usepackage{tcolorbox}

\usepackage{paralist}
\usepackage[outercaption]{sidecap}
\usepackage [autostyle, english = american]{csquotes}
\MakeOuterQuote{"}
\newcounter{scn}
\usepackage[shortlabels]{enumitem}
\usepackage{bchart}

\newcounter{finding_counter}
\newtoggle{comment}
\toggletrue{comment}

\usepackage{tcolorbox}
\newlist{RQ}{enumerate}{1}
\setlist[RQ, 1]{label = RQ \arabic*:}

\usepackage{forest}
\usepackage{soul}

\usepackage[multiple]{footmisc}
\usepackage{threeparttable} %%for table footnote
\usepackage{subcaption}
\usepackage{graphicx}

\usepackage[shortlabels]{enumitem}

\usepackage{pifont}
\usepackage{forest}

\newcommand{\circled}[1]{\tikz[baseline=(char.base)]{
    \node[shape=circle, draw=black, fill=black, text=white, inner sep=1pt] (char) {\small\strut #1};}}

\newcommand{\percentagebar}[1]{\rule{#1}{6pt}}
\usepackage{fontawesome5}

\usepackage{marvosym}

\begin{document}
%\pagenumbering{arabic}
%\pagestyle{plain}
\title{Automatic High-Level Test Case Generation using Large Language Models}

\author{Navid Bin Hasan$^{1,\dag}$, Md. Ashraful Islam$^{1,\dag}$, Junaed Younus Khan$^{1,\dag,*}$, Sanjida Senjik$^{1}$, and Anindya Iqbal$^{1}$ \\
$^{1}$Bangladesh University of Engineering and Technology \\
$^{\dag}$These authors contributed equally to this work \\
$^{*}$Corresponding Author: junaed@cse.buet.ac.bd
}

%\markboth{IEEE TRANSACTIONS ON SOFTWARE ENGINEERING,~Vol.~X, No.~X,
%Month~2018}%
%{Uddin \MakeLowercase{\textit{et al.}}: Opinion Value Analysis in
%API Reviews}

% \author{
% \IEEEauthorblockN{Junaed Younus Khan$^a$, Gias Uddin$^a$
% \\$^a$University of Calgary}
% }

%\address{$^a$Bangladesh University of Engineering and Technology and $^b$University of Calgary}

% \author{

% \IEEEauthorblockN{Junaed Younus Khan}
% \IEEEauthorblockA{\textit{Department of Computer Science and Engineering} \\
% \textit{Bangladesh University of Engineering and Technology}\\
% %Dhaka, Bangladesh \\
% %1405051.jyk@ugrad.cse.buet.ac.bd}

% \and

% \IEEEauthorblockN{Md. Tawkat Islam Khondaker}
% \IEEEauthorblockA{\textit{Department of Computer Science and Engineering} \\
% \textit{Bangladesh University of Engineering and Technology}\\
% %Dhaka, Bangladesh \\
% %1405036.mtik@ugrad.cse.buet.ac.bd}

% \and

% \IEEEauthorblockN{Gias Uddin}
% \IEEEauthorblockA{\textit{Department of Electrical and Computer Engineering} \\
% \textit{University of Calgary}\\
% %Alberta, Canada \\
% %gias.uddin@ucalgary.ca}

% \and

% \IEEEauthorblockN{Anindya Iqbal}
% \IEEEauthorblockA{\textit{Department of Computer Science and Engineering} \\
% \textit{Bangladesh University of Engineering and Technology}\\
% %Dhaka, Bangladesh \\
% %anindya@cse.buet.ac.bd}
% }

\IEEEtitleabstractindextext{%
\begin{abstract}
We explored the challenges practitioners face in software testing and proposed automated solutions to address these obstacles. We began with a survey of local software companies and 26 practitioners, revealing that the primary challenge is not writing test scripts but aligning testing efforts with business requirements. Based on these insights, we constructed a use-case $\rightarrow$ (high-level) test-cases dataset to train/fine-tune models for generating high-level test cases. High-level test cases specify what aspects of the software's functionality need to be tested, along with the expected outcomes. We evaluated large language models, such as GPT-4o, Gemini, LLaMA 3.1 8B, and Mistral 7B, where fine-tuning (the latter two) yields improved performance. A final (human evaluation) survey confirmed the effectiveness of these generated test cases. Our proactive approach strengthens requirement-testing alignment and facilitates early test case generation to streamline development.

%In software development, aligning the final product with specified requirements is crucial but challenging due to frequent misunderstandings and bugs. While the benefits of early test case generation are well recognized for addressing these challenges, our research contributes a novel method that enhances this practice. We leveraged a trained model (SOLAR-10.7B) to efficiently convert a diverse set of use cases into detailed test cases early in the development cycle. We compiled a robust dataset of 1,515 use cases to test case pairs, which facilitated the fine-tuning of our model through parameter-efficient techniques. The fine-tuned model achieved a BLEU score of 55.22\%, demonstrating its capability to generate precise and reliable test cases. \navid{Is this up to date?}

\end{abstract}

}

%
%\ccsdesc[500]{Software and its engineering~Software libraries and repositories}
%\ccsdesc[300]{Computer systems organization~Redundancy}
%\ccsdesc{Computer systems organization~Robotics}
%\ccsdesc[100]{Networks~Network reliability}

%\keywords{API, Usage Scenario, Crowd-Sourced API Documentation, Summarization}

\maketitle

\IEEEdisplaynontitleabstractindextext
\section{Introduction}\label{sec:introduction}
% %------------------------------------------

% \textit{``If you can't test it, don't build it''} -- the quote by Boris Beizer\footnote{software engineer and author, known for his contribution to system architecture and software testing} encapsulates one of the core principles of software engineering: testing is not an afterthought but a critical element that must be integrated into the design and development process.

% \textit{``If you can't test it, don't build it''}, wisely stated by Boris Beizer\footnote{software engineer and author of ``Software System Testing and Quality Assurance'' \cite{beizer1984software}}, underscores the critical role of software testing in the development lifecycle. 

Software testing is not an afterthought but a critical element that must be integrated into the design and development process \cite{everett2007software}. It ensures that the software functions as intended \cite{basili1987comparing}, meets user requirements \cite{craig2002systematic}, and reveals bugs \cite{islam2023evolution}. However, software testing faces significant challenges \cite{garousi2020exploring}. Manual testing is time-consuming \cite{bertolino2007software}, and as software requirements evolve, maintaining and updating test cases becomes increasingly complex \cite{harrold1999testing}. 

%It also checks whether security requirements and robustness are met under certain conditions \cite{potter2004software, rosero201615}. 

\begin{figure}[!t]
\centering
	\centering
   	\includegraphics[scale=.2]{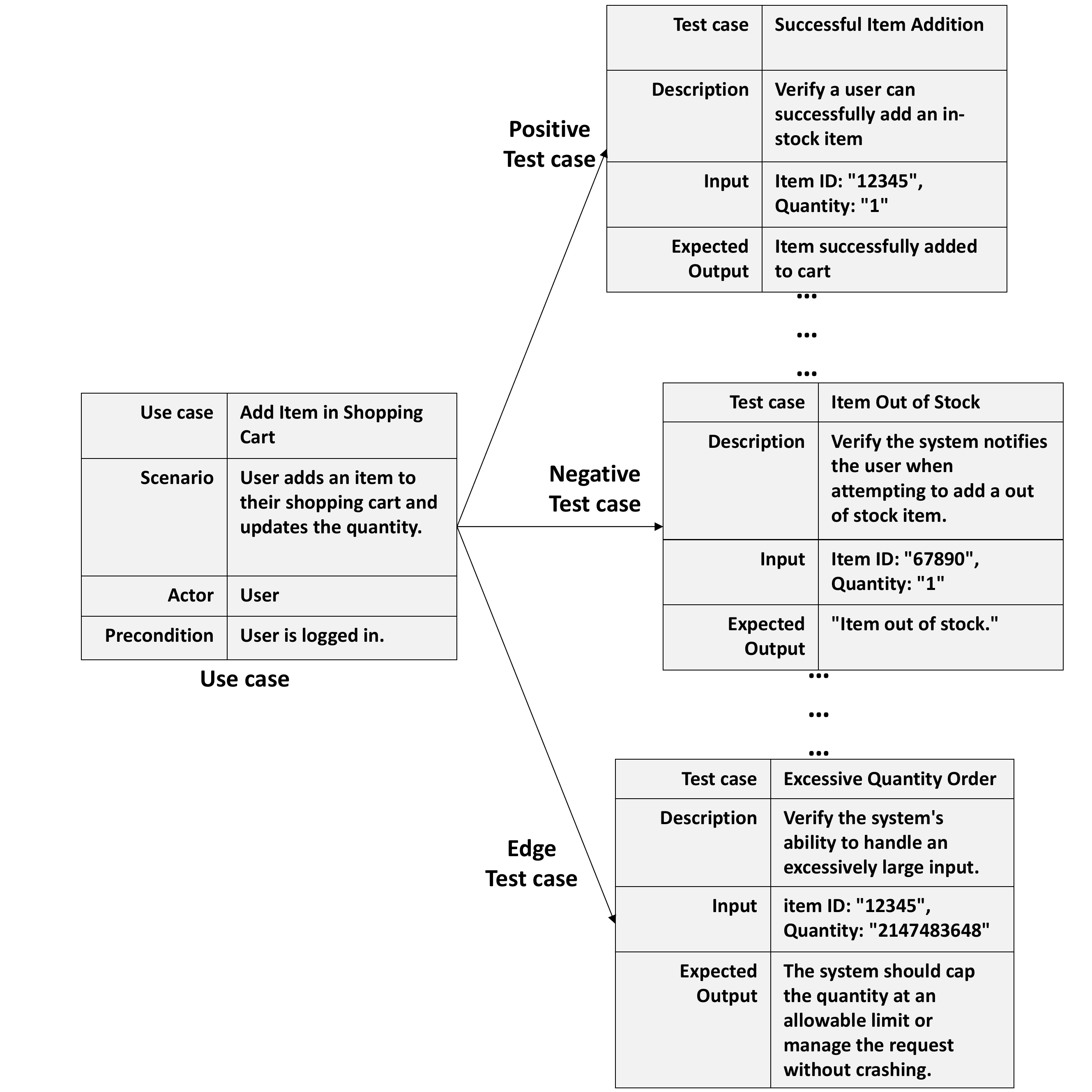}
   	\caption{Use case to Test cases Generation.}
   	 \label{fig:use_to_test_example}
\vspace{-5mm}
\end{figure}

We conducted a pilot survey by consulting three local software companies to understand their challenges in software testing. We found that the software industry is experiencing a shortage of skilled testers, partly due to testing being perceived as less prestigious than other software development activities. While there is room for improvement in the coding part of testing (i.e., writing test scripts), the main challenge lies in communicating the business requirements with the testers. Given the business requirements (e.g., use cases), most testers suffer more to understand what needs to be tested than converting them to test scripts. Specifically, testers often struggle to effectively identify and design tests for edge and negative cases, which are critical for ensuring the system can handle exceptional and out-of-bound scenarios. For example, Figure \ref{fig:use_to_test_example} depicts how a simple use case -- `Add Item in Shopping Cart' -- can be expanded into a range of test cases that address both expected and exceptional conditions.

%% Reducing
%Testing only the fundamental functionality (in this case, adding an item works as intended) is not enough. The negative test case such as `Item out of stock' is crucial for ensuring that the system handles the scenario  when an item is not available. On the other hand, edge cases like `Excessive Quantity Order' examine the system’s ability to handle abnormal inputs, such as an excessively large order quantity that could be input by error or as an attempted exploit. If not taken care of, this might be treated unexpectedly by the system. Depending on the implementation and the datatype of the corresponding variable, such overly large numbers can be interpreted as negative values due to integer overflow. This could bypass certain types of logic or validation or may even cause the system to crash.

%They serve as a guide for testers during the creation of detailed low-level test script by providing them with a clear understanding of the intended system behavior, including potential negative and edge cases.

To that end, we aim to design an automatic system that can generate all possible high-level test cases for a given use case. Unlike low-level test cases (i.e., test scripts), high-level test cases describe what needs to be tested in terms of the functionality and behavior of the software. The goal is to guide the testers during the creation of detailed low-level test script by providing them with a clear understanding of the intended system behavior, including potential negative and edge cases. 

\begin{figure}[!t]
\centering
	\centering
   	\includegraphics[scale=.6]{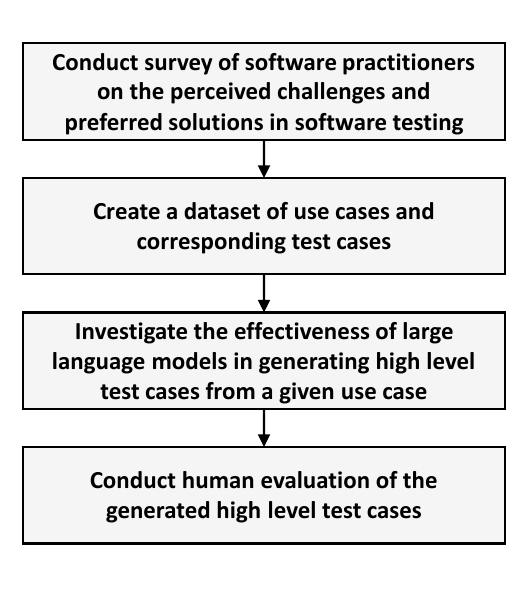}
   	\caption{The four major phases of this study.}
   	 \label{fig:workflow}
\vspace{-5mm}
\end{figure}

We conduct our study in four phases (as summarized in Figure \ref{fig:workflow}). In the initial phase, we conducted a broader survey to verify the findings and assess whether automated high-level test case generation would be a preferable solution for the practitioners. We surveyed 26 software practitioners working in different roles, companies, and countries. They confirmed that a communication gap indeed exists where ensuring the team understands what needs to be tested (based on business requirements) is a key challenge. Most of them agree that automatic high-level test case generation can be beneficial in alleviating such issues. However, they raised concerns about using external server-based tools due to company policy regarding data security and confidentiality.    

%High-level test cases focus on what needs to be tested in terms of functionality and behavior, providing a bridge between business requirements and technical testing efforts. Unlike low-level test cases that are typically written in code and executed automatically, high-level test cases are designed to ensure that testing is aligned with business goals and user needs.

% In agile software development, software companies strive for rapid response to changes in requirements. Hence, quick development and quick testing are crucial for agile. In fact, many companies nowadays adopt test-driven development (TDD) that prioritizes creating test cases before the actual code is written. Though there are mixed opinion about the feasibility and benefit of TDD, it is clear that early test case generation is beneficial for software companies that follow agile methodology.

%As found in the survey, practitioners demanded automatic solutions that can generate high-level test cases from requirement documents such as use cases. To facilitate that, in the second phase, we created a dataset of use cases and corresponding test cases for training/fine-tuning and evaluating automatic models.

In the second phase, we created a dataset of use cases and corresponding test cases to facilitate automatic generation (i.e., training/fine-tuning and evaluating machine learning models).
% \textcolor{blue}{ As we have collected sample usecase testcase pairs from different projects to be developed by the developers to build the dataset, Human Validation of the datasets were done while adding some annotations for the sake of better understanding of the of usecases by the LLMs, so that possible relevant test-cases are generated}.
The dataset contains a total of 1067 samples (i.e., 1067 use cases and their corresponding test cases) coming from diverse types of projects (such as E-commerce, Finance, EdTech, Ride-sharing, etc.). The dataset is made publicly available for future study.

In the third phase, we investigate the effectiveness of large language models in generating high-level test cases using our dataset. We evaluated the performance of different large language models (i.e., GPT-4o \cite{achiam2023gpt}, Gemini \cite{team2023gemini}, LLaMA 3.1 \cite{touvron2023LLaMA} and Mistral \cite{jiang2023mistral} ) to generate high-level test cases from use cases. While the pre-trained (raw) models show decent performance with BERTScore as high as 88.43\% (F1), fine-tuning smaller models further improves the results. We fine-tuned the LLaMA 3.1 and Mistral (comparatively smaller models) with 80\% of our dataset, and their BERTScore improved from 79.85\% to 89.94\% (for LLaMa) and from 82.86\% to 90.14\% (for Mistral). These fine-tuned smaller models present a suitable option for organizations that prioritize data confidentiality and require deployment on local servers.

% {Third, to complement the previous assessment, we used 80\% of our dataset to fine-tune the Llama 3.1 8B and Mistral 7B models and assessed their performance using the rest of the dataset. These models were chosen due to their open-source availability, strong baseline performance, fast inference times, and high efficiency. These relatively smaller models, after being fine-tuned for a specific task, are expected to replicate or surpass the performance of the larger general-purpose models.}

In the final phase, we conducted another survey to assess the quality of the generated test cases by 26 human evaluators. They evaluated the generated test cases in five different aspects (i.e., Readability, Correctness, Completeness, Relevance, Usability) on a scale from 1 to 5. The overall quality of the automated test cases, generated by the pre-trained (GPT-4o) and fine-tuned (LLaMA 3.1) models, was found to be satisfactory ($>$3.5 score in every aspect) with strong readability ($avg=4.25$), correctness ($avg=4.19$), and usability ($avg=4.32$).

We propose a proactive strategy focusing on early test case generation immediately following requirement analysis. By generating these test cases early in the development process, we can help mitigate the communication gaps that often lead to misaligned requirements and missed testing scenarios. Thus, developers can identify potential problems earlier and streamline the overall development process.

\noindent
\textbf{Replication Package.} Our code, data, and online appendix are shared at \url{https://github.com/navidh86/usecase-to-testcase}
% \url{https://anonymous.4open.science/r/uctc}

\section{Motivational Survey}
In a pilot survey with three local software companies, we found that a communication gap exists between those defining the requirements (like project managers and business analysts) and the teams responsible for testing. This often leads to software that doesn't meet specifications or receives inadequate testing. According to one CTO, ``The primary challenge in software testing is not writing test scripts but ensuring the team understands what needs to be tested based on business requirements''. This insight led us to explore the potential of high-level test cases – comprehensive outlines of testing needs that could bridge the gap between requirements and testing activities. 

% To investigate this further, we conducted a broader survey of 26 software practitioners and formulated the following research question. 

% \textbf{RQ1. How do software development teams perceive the challenges of aligning testing with business requirements?}

% We divide this question into four sub-RQs as follows.

To verify these claims, we conducted a broader survey of 26 software practitioners, where we investigated how software development teams perceive and want to address the challenges of aligning their testing efforts with business requirements. We answer the following RQ.

\begin{enumerate}[label=\textbf{RQ\arabic{*}}., leftmargin=28pt]
  \item \textbf{How do software development teams perceive their challenges in aligning testing with business requirements, and the role of automated high-level test cases in addressing these challenges?}
\end{enumerate}

% \noindent\textbf{RQ1. How do software development teams perceive their challenges in aligning testing with business requirements, and the role of automated high-level test cases in addressing these challenges?}
\noindent
We further break down this RQ into the following sub-RQs:

\begin{enumerate}[label=\textbf{RQ1.\arabic{*}.}, leftmargin=35pt]
  \item How frequently do software development teams encounter difficulties understanding what needs to be tested?
  \item What is the perceived impact of high-level test cases in guiding testing efforts?
  \item How do software practitioners perceive the automation of high-level test case generation?
  \item What are the practical concerns, such as data confidentiality and organizational policies, that might influence the adoption of a tool for automated high-level test case generation?
\end{enumerate}

\subsection{Survey Setup}
%The survey was designed to explore the challenges faced by software development teams in aligning their testing efforts with business requirements and to evaluate the potential impact of automating high-level test case generation. 

\subsubsection{Survey Questions} 
We asked each participant 8 questions listed in Table \ref{tab:survey_questions_mapping}. The questions focus on exploring the challenges faced by software development teams in aligning their testing efforts with business requirements and evaluating the potential impact of automating high-level test case generation. The survey questionnaire was designed based on our pilot survey, where we discussed the challenges of software testing with industry professionals. The recruitment of those professionals was conducted mostly by convenient sampling from the authors' professional network.

\subsubsection{Survey Participants}
The survey included responses from 26 participants across various roles, organisations, and countries. Table \ref{tab:demography_of_survey_participants} shows the distribution of the participants over their profession and experience.  

\begin{table}[h]
  \centering
  \caption{Demography of survey participants}
  \label{tab:demography_of_survey_participants}
  \resizebox{\columnwidth}{!}{%
    \begin{tabular}{rccccc}
      \textbf{} & \multicolumn{5}{c}{\textbf{Years of Experience}} \\
      \cline{2-6}
      \multicolumn{1}{r|}{\textbf{Current Role}} & \textbf{0-2} & \textbf{3-5} & \textbf{6-10} & \textbf{10+} & \textbf{Total} \\
      \hline
      \multicolumn{1}{r|}{\begin{tabular}[c]{@{}r@{}} {Developer}\end{tabular}} & 10 & 5 & 1 & - & 16 \\
      \hline
      \multicolumn{1}{r|}{\begin{tabular}[c]{@{}r@{}} {Tester}\end{tabular}} & 5 & - & 1 & 1 & 7 \\
      \hline
      \multicolumn{1}{r|}{\begin{tabular}[c]{@{}r@{}} {Team Lead}\end{tabular}} & - & 2 & - & - & 2 \\
      \hline
      \multicolumn{1}{r|}{\begin{tabular}[c]{@{}r@{}} {Project Manager}\end{tabular}} & - & 1 & - & - & 1 \\
      \hline
      \multicolumn{1}{r|}{\textbf{Total}} & \textbf{15} & \textbf{8} & \textbf{2} & \textbf{1} & \textbf{26} \\
    \end{tabular}%
  }
\vspace{-20pt}%%%%
\end{table}

\begin{table*}
\centering
\caption{Survey questions and their mapping to the Research Questions.}
\label{tab:survey_questions_mapping}
\resizebox{\textwidth}{!}
{
\begin{tabular}{c l c}
\hline
\textbf{\#} & \textbf{Questions}  & \textbf{RQ} \\ \hline
1            & How often do you or your team face difficulties in fully understanding what needs to be tested? & 1.1         \\
2            & In your experience, which aspect is more challenging for you or your team for software testing?  & 1.1         \\
3            & Do you or your team currently utilize high-level test cases? & 1.2         \\
4            & If yes, who is primarily responsible for creating high-level test cases in your organization? & 1.2         \\
5            & How useful do you believe high-level test cases are in guiding manual testing efforts? & 1.2         \\
6           & Do you or your team currently use any tool to generate or manage high-level test cases? & 1.3         \\
7           & Would you be interested in using a tool that automatically generates high-level test cases? & 1.3         \\
% 7            & To what extent would automated high-level test case generation impact test-driven development? & 1.3         \\
%8            & To what extent would automated high-level test case generation impact the traditional development? & 1.3         \\
% 10           & Do you or your team currently use any tool to generate or manage high-level test cases? & 1.3         \\
8           & Does your organization currently have policies that restrict the use of external server-based tools? & 1.4         \\

%9           & Would your organization prefer a tool that can be hosted on internal servers? & 1.4         \\ 
\hline
\end{tabular}%
}
\vspace{-4mm}
\end{table*}

\subsection{Challenges in Software Testing (RQ1.1)}
\nd\circled{Q1} We asked the participants how often they face difficulties in
understanding business requirements when determining what needs
to be tested. The responses revealed most practitioners face such issues on a frequent basis.

\begin{itemize}[leftmargin=20pt,nolistsep]
    \setlength\itemsep{0em}
    \item {\textit{\textbf{Always:}}~\color{black!90}\percentagebar{2pt}} \textbf{3.8\%}
    \item {\textit{\textbf{Often:}}~\color{black!90}\percentagebar{12pt}} \textbf{15.4\%} 
    \item {\textit{\textbf{Sometimes:}}~\color{black!90}\percentagebar{4pt}} \textbf{57.7\%}
    \item {\textit{\textbf{Rarely:}~\color{black!90}\percentagebar{6pt}}} \textbf{23.1\%}
    \item {\textit{\textbf{Never:}}~ \textbf{0\%}}
    
\end{itemize}

\nd\circled{Q2} We asked the participants about the most challenging aspect of software testing. Most participants (25) reported that the most challenging aspect of software testing is ensuring a clear understanding of what needs to be tested based on business requirements.

\begin{itemize}[leftmargin=20pt,nolistsep]
    \setlength\itemsep{0em}
    \item {\textit{\textbf{Writing test scripts:}}~\color{black!90}\percentagebar{2pt}} \textbf{3.8\%}
    \item {\textit{\textbf{Determining what to test:}}~\color{black!90}\percentagebar{25pt}} \textbf{96.2\%} 
    
\end{itemize}

\subsection{Impact of High-Level Test Cases (RQ1.2)}
\label{sec:high-level-test-case-impact}

\nd\circled{Q3} Participants were asked whether they currently use high-level
test cases or not. Most participants (22) responded in the affirmative, with a few exceptions (4).

\begin{itemize}[leftmargin=20pt,nolistsep]
    \setlength\itemsep{0em}
    \item {\textit{\textbf{Yes:}}~\color{black!90}\percentagebar{22pt}} \textbf{84.6\%}
    \item {\textit{\textbf{No:}}~\color{black!90}\percentagebar{4pt}} \textbf{15.4\%} 
    
\end{itemize}

\nd\circled{Q4} For the participants who currently use high-level test cases,
we further asked them who is primarily responsible for creating
those test cases in their organization. While the testers are mostly responsible for writing such test cases, a few exceptions (i.e., project managers, and developers) are also noted. 

\begin{itemize}[leftmargin=20pt,nolistsep]
    \setlength\itemsep{0em}
    \item {\textit{\textbf{Developer:}}~\color{black!90}\percentagebar{5pt}} \textbf{19.2\%}
    \item {\textit{\textbf{Tester:}}~\color{black!90}\percentagebar{12pt}} \textbf{46.2\%}
    \item {\textit{\textbf{Project Manager:}}~\color{black!90}\percentagebar{5pt}} \textbf{19.2\%}
    \item {\textit{\textbf{High-level test cases not used:}}~\color{black!90}\percentagebar{4pt}} \textbf{15.4\%} 
    
\end{itemize}

\nd\circled{Q5} We asked the participants about the perceived usefulness of
high-level test cases in guiding the development of test scripts
or manual testing efforts. The responses indicated a generally
positive perception of their usefulness.

\begin{itemize}[leftmargin=20pt,nolistsep]
    \setlength\itemsep{0em}
    \item {\textit{\textbf{Very Useful:}}~\color{black!90}\percentagebar{21pt}} \textbf{80.8\%}
    \item {\textit{\textbf{Moderately Useful:}}~\color{black!90}\percentagebar{5pt}} \textbf{19.2\%}
    \item {\textit{\textbf{Slightly Useful:}}~\color{black!90}\percentagebar{0pt}} \textbf{0\%}
    \item {\textit{\textbf{Not Useful:}}~\color{black!90}\percentagebar{0pt}} \textbf{0\%} 
    
\end{itemize}

\subsection{Need for Automatic High-level Test Cases (RQ1.3)}
\nd\circled{Q6} Participants were asked whether they use any tools for generating or managing high-level test cases. The majority of the participants indicated that they do not use any specific tools for high-level test cases. 

\begin{itemize}[leftmargin=20pt,nolistsep]
    \setlength\itemsep{0em}
    \item \textit{\textbf{Yes:}}~\color{black!90}\percentagebar{5pt} \textbf{19.2\%}
    \item {\textit{\textbf{No:}}~\color{black!90}\percentagebar{21pt}} \textbf{80.8\%} 
    
\end{itemize}

\noindent
Among the tools mentioned, TestRail was the most common, cited by 2 participants. One participant informed that they use a machine-learning based model (developed by themselves) for this task.

\begin{itemize}[leftmargin=20pt,nolistsep]
    \setlength\itemsep{0em}
    \item \textit{\textbf{TestRail}}~\color{black!90}\percentagebar{3pt} \textbf{7.7\%}  \item\textit{\textbf{TestCraft}}~\color{black!90}\percentagebar{2pt} \textbf{3.8\%}
     \item\textit{\textbf{Jira}}~\color{black!90}\percentagebar{2pt} \textbf{3.8\%}
     \item\textit{\textbf{ML Model}}~\color{black!90}\percentagebar{2pt} \textbf{3.8\%}
\end{itemize}

\nd\circled{Q7} We inquired about the participants' interest in using a tool that automatically generates high-level test cases based on business requirements. The responses indicated a generally positive attitude towards such a tool.

\begin{itemize}[leftmargin=20pt,nolistsep]
    \setlength\itemsep{0em}
    \item \textit{\textbf{Yes:}}~\color{black!90}\percentagebar{15pt} \textbf{57.7\%}
    \item \textit{\textbf{Maybe:}}~\color{black!90}\percentagebar{10pt} \textbf{38.5\%}
    \item {\textit{\textbf{No:}}~\color{black!90}\percentagebar{2pt}} \textbf{3.8\%} 
\end{itemize}

\subsection{Practical Concerns of Automatic Tool (RQ1.4)}
\nd\circled{Q8} Participants were asked about the existence of any organizational policies that might restrict the use of external server-based tools (for software testing). The majority responded that they have such restrictions due to confidentiality and budget issues.

\begin{itemize}[leftmargin=20pt,nolistsep]
    \setlength\itemsep{0em}
    \item \textit{\textbf{Yes:}}~{\textbf{Due to Confidentiality:}}~\color{black!90}\percentagebar{9pt} \textbf{34.6\%,}
    {\textbf{Cost:}}~\color{black!90}\percentagebar{4pt} \textbf{11.6\%}
    \item \textit{\textbf{Not sure:}}~\color{black!90}\percentagebar{9pt} \textbf{34.6\%}
    \item {\textit{\textbf{No:}}~\color{black!90}\percentagebar{5pt}} \textbf{19.2\%} 
    
\end{itemize}

\begin{tcolorbox}[
       left=0pt, right=0pt, top=0pt, bottom=0pt, colback=white, after=\ignorespacesafterend\par\noindent]
\textbf{Summary of RQ1.} Software practitioners often face challenges in understanding business requirements, with 96.2\% identifying ``determining what to test'' as the most difficult aspect of software testing. High-level test cases are widely used, with 84.6\% of participants confirming their use, and most participants (80.8\%) view them as highly useful in guiding their testing efforts. However, only 19.2\% currently use tools for maintaining/generating high-level test cases, despite the majority's willingness to adopt automated solutions if available. Confidentiality and budget concerns are major barriers, as 46.2\% of participants reported that their companies restrict the use of tools hosted on external servers.
\end{tcolorbox}

% \subsection{The Challenge of Aligning Testing with Business Requirement (RQ1)}

% \subsection{The Impact of High Level Test Cases (RQ2)}

\newcolumntype{C}[1]{>{\centering\arraybackslash}m{#1}}

\section{Automatic Generation of High-Level Test Cases}
Our survey revealed that a major challenge in software testing is that testers often struggle with understanding what needs to be tested rather than just focusing on writing test scripts. In this context, high-level test cases can provide valuable clarity. The survey participants also express interest in automated tools to generate high-level test cases. In this section, we explore the automatic generation of such test cases from business requirements documents. While we could use any form of business requirement document, we focus on use cases because i) they are very common (most software companies maintain them), and ii) they are typically written in a structured format that clearly defines the interactions between users and the system. 

% \noindent\textbf{RQ2. How effective are large language models in generating high-level test cases?}

% \noindent\textbf{RQ3. How effective are smaller models for local server deployment?}

\begin{enumerate}[label=\textbf{RQ\arabic{*}}.,start=2, leftmargin=28pt]
  \item \textbf{How effective are pre-trained large language models in generating high-level test cases?}
  \item  \textbf{How effective are smaller fine-tuned models in generating high-level test cases}?
\end{enumerate}

Below, we first describe our dataset creation process. Then, we describe the motivation and approach for each research question and report the results. 

\subsection{Dataset Creation}
To the best of our knowledge, no existing dataset provides paired use cases and their corresponding test cases. Hence, we developed a novel dataset leveraging both manually coded and synthesized data to encompass a broad spectrum of real-world and custom scenarios. Table \ref{tab:dataset_summary} provides a summary of the dataset's composition, while Table \ref{tab:project_type_stats} outlines the distribution of project types within the dataset.

\begin{table}[h]
\centering
\caption{Summary of dataset}
\label{tab:dataset_summary}
\begin{tabular}{l|l|c|c}
%\hline
\textbf{}   & \textbf{Project Type (and Count)}   & \textbf{\#Usecase} & \textbf{\#Testcase} \\ 
\hline
\multirow{2}{*}{\begin{tabular}[c]{@{}l@{}}\textbf{Human} \\ \textbf{Generated}\end{tabular}}   
                                   & Real World Projects (20) & 299 & 1229 \\
\cline{2-4}
                                   & Student Projects (42)    & 281 & 961  \\ 
\hline
\textbf{Synthesized}   & UiPath Projects (106)      & 487 & 1416 \\ 
\hline
\multicolumn{2}{r|}{\textbf{Total}}   & \textbf{1067} & \textbf{3606} \\ 
%\hline
\end{tabular}
\vspace{-15pt}
\end{table}

% \begin{table}[h]
% \centering
% \caption{Summary of dataset. The last column highlights the minimum and maximum number of test cases per use case.}
% \label{tab:dataset_summary}
% \begin{tabular}{C{1.5cm}|C{1.3cm}|C{1.1cm}|C{1.1cm}|C{1.2cm}}
% %\hline
%  \textbf{} & \textbf{Project Type (and Count)}   & \textbf{\#Usecase} & \textbf{\#Testcase} & \textbf{\#Testcases Per Usecase} \\ 
% \hline
% \multirow{2}{*}{\begin{tabular}{p{1.3cm}} \textbf{Human Generated} \end{tabular}} & Real World Projects (20) & 299 & 1229 & 1-11 \\
% \cline{2-5}
%      & Student Projects (42)    & 281 & 961 & 1-33  \\ 
% \hline
% \textbf{Synthesized}   & UiPath Projects (106) & 487 & 1416 & 1-16 \\ 
% \hline
% \multicolumn{2}{r|}{\textbf{Total}}   & \textbf{1067} & \textbf{3606} & - \\ 
% %\hline
% \end{tabular}
% \end{table}

\subsubsection{Manual Coding} The first part of the dataset was created through the collaboration of $300$ undergraduate students ($120$ third-year and $180$ fourth-year students) enrolled in advanced software development courses. The students had prior experience in software development through academic projects and internships. The students worked in groups of $5$-$6$ members, with each group supervised by (at least) one faculty member to ensure the quality of the use cases and their corresponding test cases. This effort resulted in a total of $580$ use cases and $2190$ test cases across two main categories: Real-World Applications and Student Projects.

\textbf{\underline{Real-world Applications}:} In this category, we provided students with $20$ diverse real-world applications, covering sectors such as finance, education, ride-sharing, and social media. This approach aimed to capture various practical and complex requirements reflective of industry settings. Thus we collected $299$ use cases and $1299$ test cases.

\textbf{\underline{Student Projects}:} This category comprises $281$ use cases and $961$ test cases originating from $42$ unique term projects. These student projects allowed greater flexibility and creativity that added unique perspectives to the dataset. Each project was supervised by faculty that ensures adherence to academic standards and a coherent structure within the use-case and test-case pairs.

\subsubsection{Data Synthesis} \label{sec:dataset_enhancement}
The dataset is further enhanced with synthesized data to facilitate the fine-tuning of smaller models. We utilized \href{https://www.uipath.com/}{UiPath} as a source of additional use cases. UiPath is a robotic process automation tool for large-scale end-to-end automation. It is used to automate repetitive tasks without human intervention. We conducted the data synthesis as follows. 

\textbf{\underline{UiPath Projects}:} First, we extracted $1237$ use cases from $296$ UiPath projects. Second, we synthesized test cases for the extracted use cases using GPT-4o (as done in Section \ref{sec:automatic_testcase_with_llm}). Third, we manually validated 1416 test cases corresponding to 487 use cases (of 106 UiPath projects). Only, the validated samples were incorporated in the original dataset (see Table \ref{tab:dataset_summary}) and used for fine-tuning.

% \textit{1) Use Case Collection.} We utilized \href{https://www.uipath.com/}{UiPath} as a source of additional use cases. UiPath is a robotic process automation tool for large-scale end-to-end automation. It is used to automate repetitive tasks without human intervention.
% We extracted $1237$ use cases from $296$ UiPath projects. \textit{2) Test Case Synthesis.} We synthesized test cases for the extracted use cases using GPT-4o (as done in Section \ref{sec:automatic_testcase_with_llm}). \textit{3) Manual Validation.} We manually validated 1416 test cases corresponding to 487 use cases (of 106 UiPath projects). The validated samples were incorporated in the original dataset (see Table \ref{tab:dataset_summary}) and used for fine-tuning.

\begin{table}[]
\caption{Distribution of project types in our dataset.}
\label{tab:project_type_stats}
\resizebox{\columnwidth}{!}{%
\begin{tabular}{@{}lcccc@{}}
\toprule
\textbf{Type}                & \textbf{\#Student Project} & \textbf{\#Real World} & \textbf{\#UIPath} & \textbf{Total} \\ \midrule
Vehicle Management  & 2                 & -            & 3        & \textbf{5}     \\
Travel/Entertinment & 3                 & 3            & 2        & \textbf{8}     \\
Ride-sharing        & -                 & 2            & -        & \textbf{2}     \\
Project Management  & 4                 & 2            & 14       & \textbf{20}    \\
Productivity        & -                 & 2            & -        & \textbf{2}     \\
Job Recruitment     & -                 & -            & 8        & \textbf{8}     \\
Healthcare          & 4                 & -            & 13       & \textbf{17}    \\
Hotel Management    & 1                 & 1            & -        & \textbf{2}     \\
Finance             & -                 & 2            & 14       & \textbf{16}    \\
EdTech              & 9                 & 3            & 4        & \textbf{16}    \\
E-commerce          & 8                 & 5            & 17       & \textbf{30}    \\
Agriculture         & 2                 & -            & 1        & \textbf{3}     \\
Miscellaneous       & 9                 & -            & 30       & \textbf{39}    \\

\textbf{Total}               & \textbf{42}                & \textbf{20}           & \textbf{106}       & \textbf{168}   \\ \bottomrule
\end{tabular}%
}
\vspace{-20pt}
\end{table}

\subsubsection{Quality Assurance} 
We implemented multiple quality control steps throughout the dataset creation process to ensure reliability. While supervisors provided feedback on projects during the creation of use cases and test cases, a final validation check was performed by one author to ensure consistency and correctness. For the synthesized data, we employed the GPT-4o model, which has demonstrated commendable performance in generating test cases from use cases (see Section \ref{sec:automatic_testcase_with_llm}). We also manually validated all the data points that were incorporated into our dataset.

% The dataset is further enhanced with synthesized data to facilitate the fine-tuning of smaller models (see Section \ref{sec:dataset_enhancement} for details). 

% Table \ref{tab:dataset_summary} depicts the summary of our dataset, and Table \ref{tab:project_type_stats} depicts the distribution of different project-types of it.

% \begin{table}[h]
% \centering
% \caption{Summary of dataset}
% \label{tab:dataset_summary}
% \begin{tabular}{l|l|c|c}
% %\hline
% \textbf{}   & \textbf{Project Type (and Count)}   & \textbf{\#Usecase} & \textbf{\#Testcase} \\ 
% \hline
% \multirow{2}{*}{\begin{tabular}[c]{@{}l@{}}\textbf{Human} \\ \textbf{Generated}\end{tabular}}   
%                                    & Real World Projects (20) & 299 & 1229 \\
% \cline{2-4}
%                                    & Student Projects (42)    & 281 & 961  \\ 
% \hline
% \textbf{Synthesized}   & UiPath Projects (106)      & 487 & 1416 \\ 
% \hline
% \multicolumn{2}{r|}{\textbf{Total}}   & \textbf{1067} & \textbf{3606} \\ 
% %\hline
% \end{tabular}
% \end{table}

\subsection{High-Level Test Cases with Pre-trained Large Language Models (RQ2)} 
\label{sec:automatic_testcase_with_llm}
\subsubsection{Motivation}
Pre-trained LLMs have achieved significant success across various software engineering tasks such as code generation \cite{islam2024mapcoder, zhou2023languageLATS, shinn2023reflexion}, code repair \cite{agentless, tang2024code}, and code documentation \cite{nam2024using, luo2024repoagent}. They often perform well without task-specific fine-tuning or specialized training \cite{brown2020language}. Hence, in this RQ, we investigate the inherent capability of large language models (LLMs) in generating high-level test cases. 

%Pre-trained large language models (LLMs) have achieved significant success across various software engineering tasks. Prompt engineering techniques, in particular, have led to notable improvements in tasks such as code generation \cite{islam2024mapcoder, zhou2023languageLATS, shinn2023reflexion}, code repair \cite{agentless, tang2024code}, and code documentation \cite{nam2024using, luo2024repoagent}. 

%Inspired by these advancements, we have developed a simple yet highly effective one-shot prompting approach, achieving an $88.43\%$ F1-score using BERTScore \cite{bert-score} as the evaluation metric.

\subsubsection{Approach}
In this study, we evaluated GPT-4o \cite{achiam2023gpt} and Gemini \cite{team2023gemini} on generating high-level test cases as detailed below.

\begin{figure}[h]
    \centering
    \includegraphics[width=0.99\linewidth]{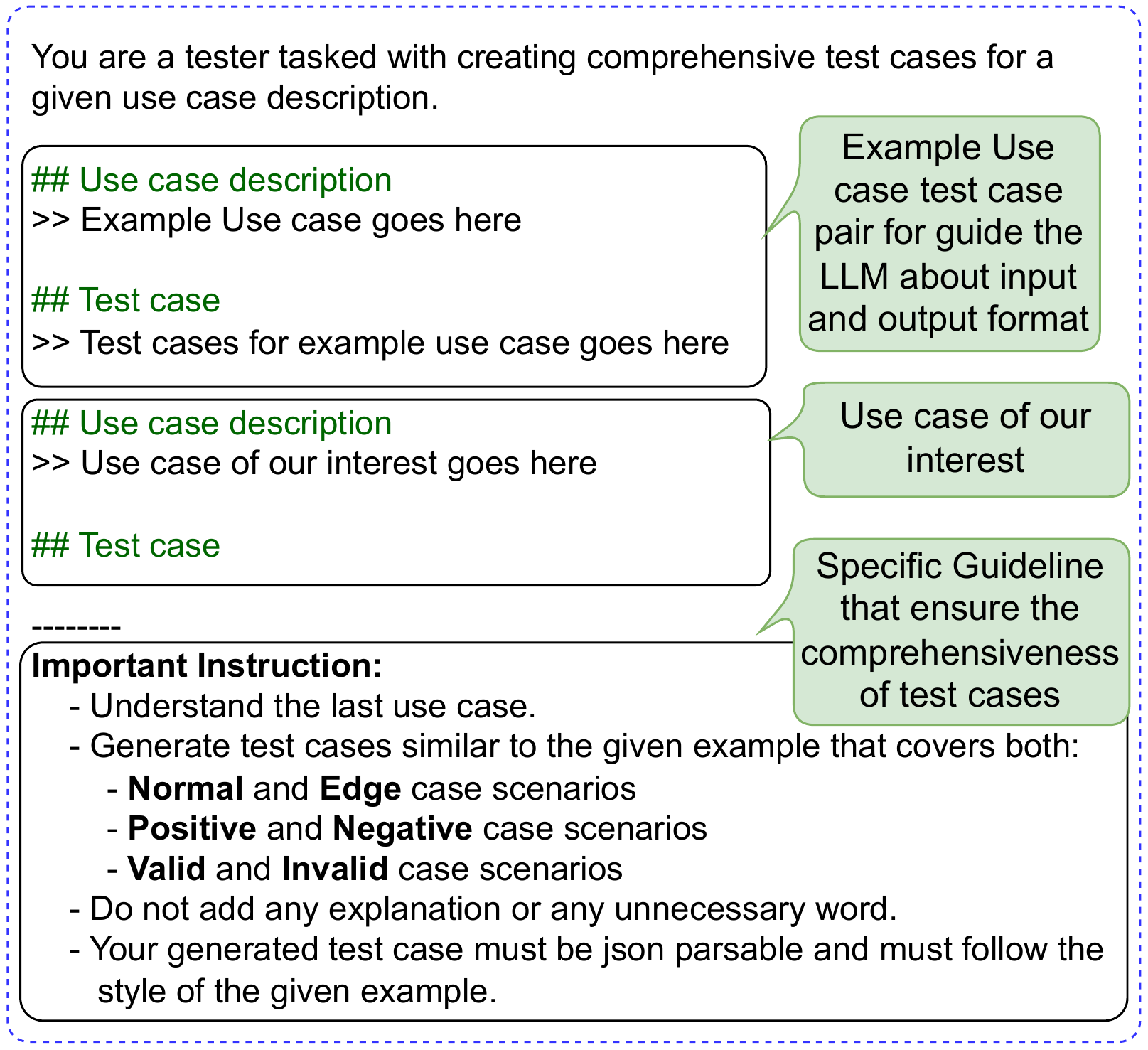}
    \vspace{0.2mm}
    \caption{Prompt for generating test cases from a use case using pre-trained LLMs.}
    \label{fig:u2t-prompt}
\end{figure}

%\junaed
{\nd\textbf{Prompt Engineering.} The interaction with an LLM (e.g., GPT-4) takes place via prompt engineering, where a task description is provided as the input (prompt), and the model performs the desired task (generates test cases) accordingly. There are several ways of prompting i.e., \textit{zero-shot, one-shot, few-shot learning} \cite{brown2020language}. In \textit{zero-shot learning}, the model is expected to generate an answer without providing any example. In fact, no additional information other than the task description itself is given in the prompt. On the other hand, one (or few)-shot learning involves giving one (or more than one) example(s) in the prompt to guide the LLMs in the correct direction. In this study, we have developed a simple yet highly effective one-shot prompting approach where we craft a generalized but well-rounded example (use case to test cases pair) that could serve as a proper guideline for the LLM, provide it in the prompt (as an example), and ask the model to generate test cases for another use case by learning from the provided example. The prompt format for generating test cases from a use case is shown in Figure \ref{fig:u2t-prompt}. The prompt begins with a single example to demonstrate the use case and test case format. Following this, the specific use case requiring test case generation is provided. Finally, the prompt includes instructions to create test cases that comprehensively cover (i) basic and edge cases, (ii) both positive and negative scenarios, and (iii) valid and invalid inputs.} The detailed prompting strategy is provided in the online appendix.

%\junaed
{\nd\textbf{Parameter Settings.} There are a number of parameters involved with large language models. One such parameter is \textit{Temperature}, which controls the randomness of the generated output (range 0 to 1). Another randomness parameter is \textit{Top-p} (range 0 to 1), which controls how unlikely words can get removed from the sampling pool by choosing from the smallest possible set (of words) whose cumulative probability exceeds \textit{p}. As recommended by OpenAI official documentation, we set the temperature at a low value (0.0) while keeping top-P at 0.95. We also keep \textit{Frequency} and \textit{Presence penalties} at their default values (0.0), which control the level of word repetition in the generated text by penalizing them based on their existing frequency and presence.}

\noindent\textbf{Evaluation Data.} For evaluation, we randomly selected a subset of $200$ human-generated data points from our dataset. This sample size provides a balance between computational efficiency and statistical validity, enabling a meaningful analysis of the model’s ability to generate test cases that align with reference outputs. 

%\textcolor{blue}{These data points were chosen from the human-generated portion of the dataset, enabling us to evaluate the model-generated test cases with the human-written ones.}

%We used the remaining portion of our dataset (i.e., $867$) for fine-tuning the model.

\noindent\textbf{Evaluation Metrics.} Several evaluation metrics exist to compare the machine-generated texts (e.g., automated test cases) with reference texts (e.g., human-generated test cases) such as ROUGE \cite{lin2004rouge}, BLEU \cite{papineni2002bleu}, BERTScore \cite{zhang2019bertscore}. Among them, {we chose BERTScore (using RoBERTa \cite{liu2019roberta} as the underlying embedding model) as our primary evaluation metric.} ROUGE (Recall-Oriented Understudy for Gisting Evaluation) measures the overlap between n-grams, longest common subsequences, and word pairs between the generated and reference texts \cite{lin2004rouge}. Conversely, the BLEU (Bilingual Evaluation Understudy) score measures the precision of n-grams between generated and reference text, often focusing on smaller n-grams such as unigrams and bigrams \cite{papineni2002bleu}. Hence, they are particularly useful for assessing surface-level or lexical similarity. Their dependence on exact matches can result in low scores even when semantically similar words or phrases are used. Unlike ROUGE and BLEU, BERTScore leverages the pre-trained contextual embeddings from transformer-based models (e.g., BERT, RoBERTa) to measure cosine similarity between the generated and reference texts. Thus, the BERTScore is less sensitive to minor lexical differences, which makes it ideal for our experiment, where the generated test cases should be semantically equivalent to the reference cases, even if they differ in wording. 
We calculated the precision, recall, and  F1 using BERTScore. Precision measures how many tokens in the generated text are semantically similar to tokens in the reference text, whereas recall measures how many tokens in the reference text are semantically similar to tokens in the generated text. The F1 score is the harmonic mean of precision and recall.

\begin{table}
\centering
\caption{Performance of pre-trained large language models.}
\label{tab:gpt-bert}
\begin{tabular}{l|c|c|c|c}
    \hline
    \textbf{Model} & \textbf{Fine-tuned} & \textbf{Precision} & \textbf{Recall} & \textbf{F1 score} \\
    \hline
    {Gemini} & \ding{55} & 86.81\% & 89.09\% & 87.92\%\\
    \hline
    {GPT-4o}& \ding{55}  & \textbf{87.63\%} & \textbf{89.27\%} & \textbf{88.43\%}\\
    \hline
\end{tabular}
\vspace{-20pt}%%%%
\end{table}

% \begin{table}[h]
% \centering
% \caption{Performance of smaller (fine-tuned) models.}
% \label{tab:finetuned-bert}
% \begin{tabular}{l|c|c|c|c}
%     \hline
%     \textbf{LLM} & \textbf{Fine-tuned} & \textbf{Precision} & \textbf{Recall} & \textbf{F1 score} \\
%     \hline
%     \textbf{LLaMA 3.1 8B} & \ding{55} & 75.7\% & 84.52\% & 79.85\% \\
%     \hline
%     \textbf{LLaMA 3.1 8B} & \checkmark & 90.11\% ↑ & 89.8\% ↑ & 89.94\% ↑ \\
%     \hline
%     \textbf{Mistral 7B} & \ding{55} & 81.36\% & 84.44\% & 82.86\% \\
%     \hline
%     \textbf{Mistral 7B} & \checkmark & 90.21\% ↑ & 90.1\% ↑ & 90.14\% ↑\\
%     \hline
% \end{tabular}
% \end{table}

\subsubsection{Result}
The results of the Gemini and GPT-4o models, as presented in Table \ref{tab:gpt-bert}, demonstrate that GPT-4o achieves higher scores across all evaluation metrics—Precision, Recall, and F1 Score—compared to Gemini. 

{GPT-4o's higher precision suggests that it is more accurate in generating relevant outputs and contains fewer irrelevant or erroneous details in the generated test cases compared to Gemini. Similarly, a higher recall for GPT-4o suggests that it retrieves more comprehensive information and is better at including all necessary details in the generated test case. Finally, the higher F1 Score for GPT-4o underscores its balanced performance in both accuracy and completeness, making it more reliable for generating test cases that closely match human-generated references.}

% \begin{enumerate}
%     \item \textbf{Precision:} GPT-4o has a Precision score of 87.63\%, showing an improvement over Gemini’s 86.81\%. This indicates that GPT-4o is more accurate in generating relevant outputs, as a higher Precision suggests fewer irrelevant or erroneous details in the generated test cases compared to Gemini.
%     \item \textbf{Recall:} GPT-4o also outperforms Gemini in Recall, with scores of 89.27\% and 89.09\%, respectively. This higher Recall implies that GPT-4o retrieves more comprehensive information and is better at including all necessary details in the generated test cases. As Recall reflects the model’s ability to capture relevant content similar to the reference cases, GPT-4o’s performance suggests a closer alignment with human-generated examples.
%     \item \textbf{F1 Score:} The F1 Score, a harmonic mean of Precision and Recall, further highlights GPT-4o’s effectiveness. GPT-4o achieves an F1 Score of 88.43\%, an increase from Gemini’s 87.92\%. The higher F1 Score for GPT-4o underscores its balanced performance in both accuracy and completeness, making it more reliable for generating test cases that closely match human-generated references.
% \end{enumerate}

% Overall, GPT-4o, with its higher Precision, Recall, and F1 Score,
% demonstrates superior capability in generating test cases that are both accurate and aligned with reference standards. 
Overall, the results suggest that GPT-4o could be more effective for tasks requiring high-quality and semantically relevant outputs, as it consistently performs better than Gemini in all aspects of evaluation.

% %For reproducibility, we have listed all the values of the hyper-parameters in Table \ref{tab:prompting-hyperparameters}.

% \begin{table}[h]
% \centering
% \caption{Hyper-parameters used for generating results of Table \ref{tab:gpt-bert}.}
% \label{tab:prompting-hyperparameters}
% \begin{tabular}{l|C{2.5cm}|C{1.5cm}|C{1cm}}
%     \hline
%     \textbf{LLM} & \textbf{Version} & \textbf{Temperature} & \textbf{Top\_p} \\
%     \hline
%     \textbf{Gemini} & gemini-1.5-pro-002 & 0 & 0.95 \\
%     \hline
%     \textbf{GPT-4o} & gpt-4o-2024-08-06 & 0 & 0.95 \\
%     \hline
% \end{tabular}
% \end{table}

% \begin{tcolorbox}[
%        left=0pt, right=0pt, top=0pt, bottom=0pt, colback=white, after=\ignorespacesafterend\par\noindent]
% \textbf{Summary of RQ2.} \junaed{Write Summary of RQ2. You can Follow the format of RQ1 summary.}
% \end{tcolorbox}

\begin{tcolorbox}[left=0pt, right=0pt, top=0pt, bottom=0pt, colback=white, after=\ignorespacesafterend\par\noindent]
\textbf{Summary of RQ2.} We evaluated two LLMs (i.e., GPT-4o and Gemini) for generating high-level test cases from given use cases. By applying a one-shot prompting strategy, we achieved promising results with GPT-4o, which surpassed Gemini in Precision, Recall, and F1 Score, as evaluated by BERT Score. This result suggests that GPT-4o, aided by prompt engineering, can produce accurate and semantically relevant test cases, making it highly effective for applications that require high-quality, contextually aligned high-level test case generation.
\end{tcolorbox}

\subsection{High-Level Test Cases with Smaller Fine-tuned Models (RQ3)}
\subsubsection{Motivation}
Many software companies are hesitant to use third-party LLMs for development tasks due to internal regulations, concerns about confidentiality, and cost. The participants of our survey also raised similar concerns. To that end, we explore the potential of smaller models that come at a cheaper price and can be deployed on the company's local server. Smaller models like LLaMA\cite{touvron2023LLaMA}, Mistral \cite{jiang2023mistral}, Gemma \cite{team2024gemma}, SOLAR \cite{kim2023solar} can be fine-tuned for a specific task with a quality dataset and can often match the performance of larger general-purpose LLMs \cite{hu2021lora,dettmers2024qlora,houlsby2019parameter,lester2021power}. Moreover, these models consume significantly less memory and computational power than larger models, allowing them to be deployed in less powerful machines. Most large-scale models like GPT-4o require a hefty price for continuous usage, whereas an open-source model deployed locally obviates the usage cost. Furthermore, an in-house LLM ensures the privacy and security of confidential data.  

%\cite{sanh2019distilbert}

% Apart from using GPT-4o, we also fine-tuned the LLaMA 3.1 8B model to automatically generate quality test cases from a given use case. 

\subsubsection{Approach}
We fine-tuned the LLaMA 3.1 8B \cite{touvron2023LLaMA} and Mistral 7B \cite{jiang2023mistral} to generate quality test cases from a given use case automatically. We chose LLaMA 3.1 8B as it can handle complex language tasks despite its smaller size and fast inference. We also fine-tuned Mistral 7B, as it has been shown to outperform the LLaMA 2 13B model on all benchmarks and even the LLaMA 134B model on some benchmarks \cite{jiang2023mistral}. It also offers practical advantages like faster inference time and handling longer sequences at a smaller cost. To assess the impact of fine-tuning, we also evaluated the raw, pre-trained models (before fine-tuning) as a baseline.

\noindent\textbf{Prompt Engineering.} We used the same prompt format outlined in Section \ref{sec:automatic_testcase_with_llm} for the fine-tuned models.

\noindent\textbf{Parameter Settings.} For fine-tuning the models, we utilized QLoRA \cite{dettmers2024qlora} with $4$-bit precision to efficiently reduce memory usage while retaining model performance. We used the \href{https://unsloth.ai/}{Unsloth} library, which promises up to $5$ times faster fine-tuning with a $70\%$ reduction in memory usage for both LLaMA and Mistral models. For parameter-efficient fine-tuning (PEFT) \cite{houlsby2019parameter} with LoRA \cite{hu2021lora}, we set $rank = 16$ and $alpha = 16$; this decreases the computational and memory cost of LoRA by storing less information. Such parameter-efficient fine-tuning on high-quality datasets often achieves high performance (comparable to state-of-the-art) while only requiring a fraction of the parameters needed for full fine-tuning \cite{hu2021lora,dettmers2024qlora,houlsby2019parameter,lester2021power}. We targeted every linear module to maximize quality. For faster training, dropout and biases were not used. The training was run for $10$ epochs with a weight decay of $0.1$ and a warm-up ratio of $0.05$.

\noindent\textbf{Evaluation Data.} We used $80\%$ of the data for fine-tuning and the remaining $20\%$ for testing. To be specific, we chose the same $200$ data points from Section \ref{sec:automatic_testcase_with_llm} as our test data (to facilitate a fair performance comparison between the two models). We used the remaining portion (i.e., $867$ data points) for fine-tuning the models. While selecting the $200$ data points (in Section \ref{sec:automatic_testcase_with_llm}), we ensured a project-wise split, which means there was no overlap of projects between the fine-tuning and test sets. Thus, the model is evaluated on projects it has not seen during fine-tuning, allowing for a realistic assessment.

\noindent\textbf{Evaluation Metrics.} Similar to Section \ref{sec:automatic_testcase_with_llm}, we used the BERTscore to evaluate the generated test cases.

\subsubsection{Result}
The results for the fine-tuned models have been reported in Table \ref{tab:finetuned-bert}. A significant boost in performance through fine-tuning can be observed. In all three metrics (precision, recall, F1), the fine-tuned models obtain scores in the range of 89-90\%, beating all the non-fine-tuned models, including GPT-4o and Gemini (Table \ref{tab:gpt-bert}). This further validates our expectation of replicating or surpassing the performance of huge models like GPT-4o and Gemini using a much smaller model fine-tuned with a quality dataset \cite{hu2021lora,dettmers2024qlora,houlsby2019parameter,lester2021power}.

\begin{table}[h]
\centering
\caption{Performance of smaller (fine-tuned) models.}
\label{tab:finetuned-bert}
\begin{tabular}{l|c|c|c|c}
    \hline
    \textbf{Model} & \textbf{Fine-tuned} & \textbf{Precision} & \textbf{Recall} & \textbf{F1 score} \\
    \hline
    {LLaMA 3.1 8B} & \ding{55} & 75.70\% & 84.52\% & 79.85\% \\
    \hline
    {LLaMA 3.1 8B} & $\boldsymbol{\checkmark}$ & \textbf{90.11\%} & \textbf{89.8\%} & \textbf{89.94\%} \\
    \hline
    {Mistral 7B} & \ding{55} & 81.36\% & 84.44\% & 82.86\% \\
    \hline
    {Mistral 7B} & $\boldsymbol{\checkmark}$ & \textbf{90.21\%}  & \textbf{90.1\%}  & \textbf{90.14\%} \\
    \hline
\end{tabular}
\end{table}

\begin{tcolorbox}[
       left=0pt, right=0pt, top=0pt, bottom=0pt, colback=white, after=\ignorespacesafterend\par\noindent]
\textbf{Summary of RQ3.} Two state-of-the-art open-source models were chosen within the 10B parameter range for fine-tuning, namely LLaMA 3.1 8B and Mistral 7B. We utilized QLoRA with $4$-bit precision to fine-tune these models using our limited hardware.  Despite the relatively smaller size of these models, they gained significant results after fine-tuning, outperforming massive pre-trained models like GPT-4o and Gemini. These results suggest that software companies can maintain in-house fine-tuned LLMs capable of generating quality test cases while ensuring privacy.
\end{tcolorbox}

\subsection{Human Evaluation}
\label{sec:human_eval}
\subsubsection{Motivation} 
While automatic metrics provide a standardized way to assess generated test cases, they may lack the nuance and contextual understanding required for a comprehensive evaluation. BERTScore, which works based on semantic similarity, can overlook critical aspects such as logical structure, the presence of edge cases, and specific testing contexts. Two test cases that appear similar in language might differ in the logical sequence of actions or conditions. BERTScore might rate them highly similar, overlooking structural discrepancies that a human evaluator would consider significant. Moreover, BERTscore doesn’t inherently capture the adequacy of edge or negative cases in generated test cases. Hence, a generated set can have high similarity to the reference but lacks critical edge or negative scenarios -- something a human evaluator would prioritize. Hence, a complete human evaluation of the automated test cases is necessary. 

%Overall, BERTScore’s effectiveness relies on pre-trained embeddings, which might lack specificity in software testing language.

\subsubsection{Approach}
We conducted a human evaluation survey to complement the limitations of BERTScore in assessing generated test cases. As mentioned in the earlier sub-sections, we evaluated the automatic models on $200$ test samples (i.e., use case and test cases pairs). At $90\%$ confidence level and $10\%$ margin of error, a statistically significant sample size would be 51. However, to facilitate efficient distribution among evaluators, we randomly selected 52 use cases, each with corresponding human-generated, GPT-4o-generated, and LLaMA 3.1-generated test cases. These samples were divided into $13$ sets (each containing $4$), and distributed to $26$ human evaluators (with each set to be reviewed by two evaluators). While there is partial overlap with the previous group of $26$ survey participants, the majority are different individuals. Among them, the majority (16) are industry professionals serving as developers or testers, 8 are academics teaching software engineering courses, and 2 are graduate students specializing in software engineering. To prevent bias, the order of the test cases was randomized, and there was no indication of whether the test cases were human-generated or machine-generated.

%%
% For a given use case, we show the test cases generated by human, GPT-4o, LLaMA (in random order) to a participant. 

The participants were asked to evaluate each set of test cases based on several criteria that are essential to high-quality software testing: Readability (clarity of reading), Correctness (functional accuracy), Completeness (coverage of scenarios), Relevance (focus on essentials), and Usability (ease of transformation into test scripts). Evaluators rated these five aspects on a Likert scale of 1 to 5 (1 being the lowest, 5 being the highest). The questionnaire for the evaluation survey is detailed in Table \ref{tab:evaluation_criteria}. 

We collected responses from all evaluators and calculated the average scores across all five aspects for each sample type (human-generated, GPT-generated, and LLaMA-generated). We measured the average interrater agreement (between the two evaluators of each set) using Cohen's Kappa \cite{Cohen-Kappa-EducationalPsy1960}. To be specific, we used quadratically weighted Cohen's Kappa as it accounts for the ordinal nature of the Likert scale (i.e., 1 to 5) and penalizes larger discrepancies between ratings more heavily than smaller ones \cite{vanbelle2016new}. The overall Cohen's Kappa between the two sets of raters was measured to be 0.42, indicating a moderate level of agreement \cite{mchugh2012interrater}.

\begin{table}[h]
\caption{Evaluation Criteria and Survey Questions}
\label{tab:evaluation_criteria}
\resizebox{\columnwidth}{!}{%
\begin{tabular}{@{}ll@{}}
\toprule
\textbf{Aspect}       & \textbf{Question}                                                                                                                                                   \\ \midrule
Readability  & \begin{tabular}[c]{@{}l@{}}How clear and easy to understand are the \\ test cases?\end{tabular}                                                            \\ \midrule
Correctness  & \begin{tabular}[c]{@{}l@{}}How accurately do the test cases capture \\ the intended functionality of the use case?\end{tabular}                            \\ \midrule
Completeness & \begin{tabular}[c]{@{}l@{}}How thoroughly do the test cases cover all \\ relevant scenarios, including positive, edge, \\ and negative cases?\end{tabular} \\ \midrule
Relevance    & \begin{tabular}[c]{@{}l@{}}How well do the test cases focus on essential \\ aspects, avoiding unnecessary or irrelevant \\ steps?\end{tabular}             \\ \midrule
Usability    & \begin{tabular}[c]{@{}l@{}}How easily can the test cases be transformed\\ into executable test scripts?\end{tabular}                                       \\ \bottomrule
\end{tabular}%
}
\end{table}

\subsubsection{Result}
The result of the human evaluation is summarized in Table \ref{tab:human_eval} and is discussed below.

\noindent\textbf{\underline{Readability}:} Both GPT-4o and LLaMA 3.1 produced clear and easy-to-understand test cases with readability scores of 4.51 and 4.54, respectively. Interestingly, these models even slightly surpassed the readability of human-generated cases, which scored 4.46. This speaks to the inherent language processing capabilities of LLMs, which are pre-trained on extensive datasets to handle a variety of linguistic patterns with fluency.

\noindent\textbf{\underline{Correctness}:} Human-generated test cases achieved the highest correctness score of 4.31. GPT-4o followed closely with a score of 4.23, while LLaMA scored lower at 4.15. This implies that GPT-generated test cases are more effective than those produced by LLaMA in capturing core functionalities accurately.

\noindent\textbf{\underline{Completeness}:} Human-generated test cases scored the highest for completeness (4.04), while GPT-generated cases had a slightly lower score (3.91), and LLaMA scored the lowest (3.61). It is interesting to see how all the samples, including the human-generated ones, suffer the most in completeness, i.e., covering all possible scenarios including edge cases and negative cases. The gap is even larger for the automated test cases.

\noindent\textbf{\underline{Relevance}:} LLaMA scored marginally higher at 4.02 on relevance, just surpassing GPT’s 3.98. Nevertheless, both models fell short of the human-generated score of 4.30. This difference suggests that while machine-generated cases succeed in capturing core functionality, they may occasionally include steps that are irrelevant. LLaMA's slightly better performance (compared to GPT) can be attributed to task-specific fine-tuning.

\noindent\textbf{\underline{Usability}:} Human-generated test cases were rated the most usable (4.43), while both GPT-4o and LLaMA followed closely with scores of 4.33 and 4.30, respectively. That suggests that given the high-level test cases, testers can easily convert them to the low-level test cases (i.e., test scripts).

% \begin{table}[]
% \caption{Human Evaluation of the Generated Test cases}
% \label{tab:human_eval}
% \resizebox{\columnwidth}{!}{%
% \begin{tabular}{lccccc}
% \hline
% \textbf{Generated By} & \textbf{Readability} & \textbf{Correctness} & \textbf{Completeness} & \textbf{Relevance} & \textbf{Usability} \\ \hline
% Human        & 4.46        & 4.31        & 4.04         & 4.30      & 4.43      \\ \hline
% GPT-4o          & 4.51        & 4.23        & 3.91         & 3.98      & 4.33      \\ \hline
% LLaMA        & 4.54        & 4.15        & 3.61         & 4.02      & 4.30      \\ \hline
% \end{tabular}%
% }
% \end{table}

\begin{table}[h]
\centering
\caption{Human evaluation of the generated test cases.}
\label{tab:human_eval}
\scalebox{1.2}{
\begin{tabular}{lccc}
\hline
\textbf{Aspect}       & \textbf{Human} & \textbf{GPT-4o} & \textbf{LLaMA 3.1} \\ \hline
{Readability}  & 4.46           & 4.51         & 4.54           \\ \hline
{Correctness}  & 4.31           & 4.23         & 4.15           \\ \hline
{Completeness} & 4.04           & 3.91         & 3.61           \\ \hline
{Relevance}    & 4.30           & 3.98         & 4.02           \\ \hline
{Usability}    & 4.43           & 4.33         & 4.30           \\ \hline
\end{tabular}%
}
\end{table}
\vspace{-15pt}

\begin{tcolorbox}[
       left=0pt, right=0pt, top=0pt, bottom=0pt, colback=white, after=\ignorespacesafterend\par\noindent]
\textbf{Summary of Human Evaluation.} The overall performance of the automated test cases is satisfactory, achieving at least moderate scores ($>3.5$) across all evaluation criteria. Both GPT-4o and LLaMA 3.1 test cases demonstrate strong readability, correctness, and usability, though they show limitations in completeness and relevance. A negative correlation is observed between completeness and relevance: GPT-4o compromises dearly on relevance to achieve completeness, often including unnecessary details, while LLaMA prioritizes relevance, which might lead to incomplete scenario coverage. This can be attributed to the inherent design of each model, where GPT’s broader training encourages extensive detail, while LLaMA’s fine-tuning emphasizes conciseness.

\end{tcolorbox}

% \begin{table}[]
% \caption{}
% \label{tab:human_eval}
% \resizebox{\columnwidth}{!}{%
% \begin{tabular}{l|c|c|c|c|c}
% \hline
% \textbf{Generated By} & \textbf{Readability} & \textbf{Correctness} & \textbf{Completeness} & \textbf{Relevance} & \textbf{Usability} \\ \hline
% Human        & 4.46        & 4.31        & 4.04         & 4.30      & 4.43      \\ \hline
% GPT-4o          & 4.51        & 4.23        & 3.91         & 3.98      & 4.33      \\ \hline
% LLaMA        & 4.54        & 4.15        & 3.61         & 4.02      & 4.30      \\ \hline
% \end{tabular}%
% }
% \end{table}

\section{Discussion}

\subsection{Impact of Enhanced Context}
\label{sec:enhanced-context}
To evaluate whether the LLMs perform better with more context, we assessed the two following approaches: 1) Providing more information about the project in the prompt with a brief overview, and 2) Using RAG to select more relevant examples for one-shot prompting.

\subsubsection{Impact of Prompts with Project Descriptions}
\label{sec:description-prompt}
Thus far, the LLMs, both pre-trained like GPT-4o and fine-tuned like LLaMA 3.1, were being instructed with one-shot prompts where they only had a use case to generate test cases from. To add more context to our prompts, we enhanced them with a brief description of the project/module/submodule to which the use case in consideration belongs. The template of the enhanced prompt has been provided in the online appendix.

% %%% Reducing
% The improved prompt is shown in Figure \ref{fig:u2t-prompt-with-description}.

% \begin{figure}[h]
%     \centering
%     \includegraphics[width=0.99\linewidth]{new_images/u2t-prompt-with-description.pdf}
%     \vspace{0.2mm}
%     \caption{Prompt for generating test cases from a use case using pre-trained LLMs.}
%     \label{fig:u2t-prompt-with-description}
% \vspace{-12pt}%%%%
% \end{figure}
% %%% Reducing

We randomly choose a subset of $100$ data points from our previous test dataset of $200$. All data points of this subset belong to the real-world project section of our dataset. We then crafted enhanced prompts for them by inserting a brief overview of their corresponding project/module.

We then evaluated the performance of GPT-4o and the fine-tuned LLaMA 3.1 8B and Mistral 7B models on the subset of the test dataset using the enhanced prompts. 

\begin{table}[h]
\centering
\caption{Performance of models with enhanced prompts using project descriptions.}
\label{tab:enhanced-bert-1}
\begin{tabular}{l|c|c|c|c}
    \hline
    \textbf{Model} & \textbf{Description} & \textbf{Precision} & \textbf{Recall} & \textbf{F1 Score} \\
    \hline
    {GPT-4o} & \ding{55} & 87.29\% & 88.77\% & 88.01\% \\
    \hline
    {GPT-4o} & $\boldsymbol{\checkmark}$ & 87.08\% & 88.87\% & 87.95\% \\
    \hline
    {LLaMA 3.1 8B} & \ding{55} & 89.30\% & 89.14\% & 89.21\% \\
    \hline
    {LLaMA 3.1 8B} & $\boldsymbol{\checkmark}$ & 89.31\% & 89.23\% & 89.26\% \\
    \hline
    {Mistral 7B} & \ding{55} & 89.40\% & 89.34\% & 89.35\% \\
    \hline
    {Mistral 7B} & $\boldsymbol{\checkmark}$ & 89.49\% & 89.14\% & 89.31\% \\
    \hline
\end{tabular}
\end{table}
\vspace{-8pt}
% \begin{table}[h]
% \centering
% \caption{BERTScores of the GPT-4o and LLaMA 3.1 8B models with enhanced context-aware prompts using descriptions.}
% \label{tab:enhanced-bert}
% \begin{tabular}{l|C{1.4cm}|C{1.4cm}|C{1.5cm}|C{1.5cm}}
%     \hline
%      & \textbf{GPT-4o (Original Prompt)} & \textbf{GPT-4o (Enhanced Prompt)} & \textbf{LLaMA 3.1 8B (Original Prompt)} & \textbf{LLaMA 3.1 8B (Enhanced Prompt)} \\
%     \hline
%     \textbf{Precision} & 87.29\% & 87.08\% & 89.30\% & 89.31\% \\
%     \hline
%     \textbf{Recall} & 88.77\% & 88.87\% & 89.14\% & 89.23\% \\
%     \hline
%     \textbf{F1} & 88.01\% & 87.95\% & 89.21\% & 89.26\% \\
%     \hline
% \end{tabular}
% \end{table}

From Table \ref{tab:enhanced-bert-1}, it is evident that the enhanced prompts do not make any significant difference in the performance of the models. This could be attributed to two possible reasons: 1) the use cases written by our human annotators are already self-sufficient, and the addition of a project description does not contribute much to the context, and 2) LLaMA 3.1 8B and Mistral 7B have already been fine-tuned without the enhanced prompts; as a result, they might not require the additional context to generate quality test cases.

%In both Figure \ref{fig:u2t-prompt} and Figure \ref{fig:u2t-prompt-with-description}, we used a consistent example to clarify the input and output format for the LLM.

\subsubsection{Impact of Using RAG}
Studies find that relevant examples given in the prompt can improve LLM performance \cite{gao2023retrieval, islam2024mapcoder}. Hence, we integrated Retrieval Augmented Generation (RAG) \cite{lewis2020retrieval} utilizing \href{https://ai.google.dev/gemini-api/docs/embeddings}{Gemini Embedding} and \href{https://pypi.org/project/chromadb/}{Chroma-DB} as our document store. Our knowledge base consists of use case–test case pairs from the student project section of our human-generated dataset. Through RAG, we retrieve the most relevant use cases from this knowledge base, which we then provide as examples in our one-shot learning prompt.

We examined the impact of incorporating RAG on the same subset of $100$ data points as used in Section \ref{sec:description-prompt}, with results summarized in Table \ref{tab:enhanced-bert-2}. Interestingly, performance gains were not as substantial as anticipated. This is because our use case–test case pairs are mostly distinct, as they come from different projects. Consequently, the retrieved examples may not be as relevant, offering limited support to the LLM and resulting in nearly equivalent performance.

\begin{table}[h]
\centering
\caption{Performance of models with enhanced prompts using RAG.}
\label{tab:enhanced-bert-2}
\begin{tabular}{l|c|c|c|c}
    \hline
    \textbf{Model} & \textbf{RAG} & \textbf{Precision} & \textbf{Recall} & \textbf{F1 Score} \\
    \hline
    {GPT-4o} & \ding{55} & 87.29\% & 88.77\% & 88.01\% \\
    \hline
    {GPT-4o} & $\boldsymbol{\checkmark}$ & 88.22\% & 88.78\% & 88.49\% \\
    \hline
    {LLaMA 3.1 8B} & \ding{55} & 89.30\% & 89.14\% & 89.21\% \\
    \hline
    {LLaMA 3.1 8B} & $\boldsymbol{\checkmark}$ & 88.91\% & 88.70\% & 88.79\% \\
    \hline
    {Mistral 7B} & \ding{55} & 89.40\% & 89.34\% & 89.35\% \\
    \hline
    {Mistral 7B} & $\boldsymbol{\checkmark}$ & 89.13\% & 88.65\% & 88.88\% \\
    \hline
\end{tabular}
\end{table}
\vspace{-15pt}

% In all the one-shot prompts we have used so far, we provided a fixed but carefully crafted generic usecase-to-testcase example to help the LLMs understand the task better.

% In this section, we explored the effects of using a more relevant example in the one-shot prompt instead of the generic one. We once again chose the same subset of 100 data points used in Section \ref{sec:description-prompt}. To generate relevant examples, we used Retrieval Augmented Generation (RAG) \navid{need citation}. 

% For our RAG pipeline, we used the usecase-testcase pairs from student project sections of our human-generated dataset as our knowledge base. Note that these are disjoint from the 100 data points we chose from the real world projects for evaluation. We used the Gemini embedding model to embed the knowledge base and Chroma-DB as the vector storage.

% \begin{table}[h]
% \centering
% \caption{BERTScores of the GPT-4o and LLaMA 3.1 8B models with enhanced one-shot prompts using RAG.}
% \label{tab:enhanced-bert}
% \begin{tabular}{l|C{1.4cm}|C{1.4cm}|C{1.5cm}|C{1.5cm}}
%     \hline
%      & \textbf{GPT-4o (Original Prompt)} & \textbf{GPT-4o (Enhanced Prompt)} & \textbf{LLaMA 3.1 8B (Original Prompt)} & \textbf{LLaMA 3.1 8B (Enhanced Prompt)} \\
%     \hline
%     \textbf{Precision} & 87.29\% & 88.22\% & 89.30\% & 88.91\% \\
%     \hline
%     \textbf{Recall} & 88.77\% & 88.78\% & 89.14\% & 88.70\% \\
%     \hline
%     \textbf{F1} & 88.01\% & 88.49\% & 89.21\% & 88.79\% \\
%     \hline
% \end{tabular}
% \end{table}

\subsection{Common Issues of LLM-Generated Test Cases}
We manually analyzed the generated test cases and identified some common issues.

% \noindent\textbf{\underline{Irrelevant Test Cases}:} LLMs sometimes generate test cases that do not align with the specified use case making assumptions not supported by the scenario. This is especially apparent in the ``Relevance'' aspect in our human evaluation (Section \ref{sec:human_eval}) and we can see how both GPT and LLaMA suffer at this (Table \ref{tab:human_eval}. This can be attributed to LLM's hallucination, a commonly known phenomenon where the model generates plausible-sounding but incorrect information. For example, for the use case ``Organize Content into Columns", (where a user arranges content side-by-side by creating multiple columns on \hyperlink{https://www.notion.so/}{Notion}, a note-taking application), GPT-4 generates a test case ``Failed Content Organization Due to Invalid Column Number" to check a scenario where users might attempt an invalid column count. But such a scenario is not relevant in the context of Notion.\\

\noindent\textbf{\underline{Irrelevant Test Cases}:} LLMs sometimes generate test cases that do not align with the specified use case making assumptions not supported by the scenario. This is especially apparent in the ``Relevance'' aspect in our human evaluation (Section \ref{sec:human_eval}) and we can see how both GPT-4o and LLaMA suffer at this (Table \ref{tab:human_eval}). This can be attributed to LLM's hallucination, a commonly known phenomenon where the model generates plausible-sounding but incorrect information. For example, for the use case ``Organize Content into Columns", (where a user arranges content side-by-side by creating multiple columns on a note-taking application), GPT-4o generates a test case ``Failed Content Organization Due to Invalid Column Number" to check a scenario where users might attempt an invalid column count. But such a scenario is not relevant in the given app's context.

\noindent\textbf{\underline{Redundant Test Cases}:} LLM can produce test cases that are essentially duplicates (with slight modifications). They test similar scenarios that do not provide additional insight. For example, for the use case ``Switch Between Delivery and Pickup While Browsing the Restaurant Menu'' of an online food ordering app, LLaMA produces the following successful cases: i) Switch to Delivery, ii) Switch to Pickup, iii) Switch to Delivery from Pickup, iv) Switch to Pickup from Delivery. These test cases are intended to check successful transitions from the `delivery' option to `pickup' option and vice-versa. As such, the first and third test cases are the same and so are the second and fourth ones.

\noindent\textbf{\underline{Insufficient Test Cases}:} LLMs may overlook test cases that human testers would typically consider, specially various edge cases. For example, for the use case ``Add Icons and Cover Art'' of a media management system, it is important to test that unsupported file-types are denied by the system. This edge case is incorporated in the human-generated test cases as ``Invalid Cover Image type'' where the test is performed with a `pdf' file instead of an image (e.g., `jpg', `png'). However, both GPT-4o and LLaMA fail to include such edge cases. 

\noindent\textbf{\underline{Lack of Input Specification}:} LLM-generated test cases can sometimes lack a structured input format, which leads to inconsistent levels of detail and clarity compared to human-generated test cases. For example, in the use case ``Add comments to Page'' of a custom social media application, the human coder adds the following input fields for a given test case: [`user', `commentText', `mentionedPersons', `mentionedGroups']. However, due to lack of proper context, the LLM-generated test cases fail to capture the `mentionedGroups' field and only include the input fields: [`user', `commentText', `mention']. Moreover, human coders are often more capable utilizing realistic data entries for input fields compared to LLMS. While such realistic input values may seem unnecessary in the context of high-level test cases (specially when using dummy data), contextually appropriate entries can be helpful in certain cases. For example, in a mapping application, realistic place names and locations enable testers to understand the functionalities under test more accurately. They provide a meaningful reference for expected tests and results.

\subsection{Implications of Study}
The implications of this study are tied to the challenges practitioners face in aligning software testing with business requirements, as highlighted in our motivational survey, where 96.2\% of respondents identified ``determining what to test” as a major issue.

User stories, which are concise narratives of system functionality from the user’s perspective, and use cases, which describe specific interactions between users and the system, both serve as critical bridges between business requirements and implementation. For example, the user story of adding items to the shopping cart in an e-commerce system can be: "As a user, I want to add items to my shopping cart so that I can purchase them later". The corresponding use case would be: “The user adds an item to their shopping cart, and the system updates the quantity”. While such user-stories/use-cases capture the functional goal, they lack the specific details necessary for identifying comprehensive testing scenarios, such as: What happens if an item is out of stock? How does the system handle invalid quantities (\textit{e.g., 0 or a very large number})? Should there be a limit on the number of items a user can add to the cart? Without these details, a (novice or less experienced) tester might only produce a simple test case: Verify that an in-stock item can be added to the cart successfully. While skilled testers could infer additional scenarios, such as handling out-of-stock items or invalid quantities, our motivational survey revealed a shortage of skilled testers in the industry. This shortage, combined with the time pressures faced by testing teams, often leads to gaps in testing coverage, even for those projects whose business requirements are well-documented (through user stories or use cases). LLMs address these challenges by generating high-level test cases, which focus on capturing comprehensive testing needs based on functional requirements. As per our findings, LLMs, trained on diverse datasets of use cases and test cases, can replicate this capability at scale. They systematically extract implicit testing needs, cover positive, negative, and edge-case scenarios, and translate high-level use-cases into actionable test cases. This reduces the cognitive load on testers and ensures alignment with business goals. Thus, LLM-generated cases can act as a foundation both for novice and experienced testers, which allows them to review, use, and refine the tests for specific project needs. Overall, it will reduce the time and resource requirements for testing.

Additionally, our approach facilitates Test-Driven Development (TDD) by enabling teams to define tests early in the development process. Early and extensive test case generation allows teams to proactively identify potential issues during the initial stages of development, improve efficiency, and reduce costly errors later in the software life cycle.

% Importantly, high-level test cases can also serve as a guide for developers during implementation by clarifying what behaviors the system is expected to support. Furthermore, managerial roles can use high-level test cases as traceable artifacts to monitor whether all business requirements are adequately developed and tested that ensures accountability and coverage throughout the project life cycle.}

\subsection{Threats to Validity}

\textit{\textbf{Internal validity}} concerns the potential inconsistencies within the study. One threat is response bias, where participants may provide answers to correspond with social desirability or perceived expectations. To address this, we assured participants of the anonymity and confidentiality of their responses. Another threat can be related to hyperparameter optimization. The models used in our study depend on hyperparameters that can significantly influence performance. Given the vast search space, exhaustive hyperparameter tuning was resource-intensive and beyond our primary objective. While the current configuration achieves satisfactory results, further optimization could yield better results.

\textit{\textbf{External validity}} relates to the generalizability of our findings to broader contexts.  The participants of both the pilot and the main survey are selected using the principle of convenience sampling from the professional network of the authors. This could incur a selection bias. In order to avoid this, we included a diverse set of practitioners across different roles, companies, countries, and experience levels. Though we selected real-world applications from diverse sectors (e.g., finance, education, social media) and incorporated original student projects to capture a wide range of testing scenarios, our dataset may not fully represent all industry scenarios.

\textit{\textbf{Construct validity}} addresses the extent to which the study measures what it intends to measure. A primary construct validity threat arises from the possibility that our survey responses do not fully capture the practical challenges software teams face in aligning testing with business requirements. We mitigated this by designing specific survey questions based on pilot study findings and mapping them directly to our research questions, ensuring alignment with our study objectives. The second concern regarding construct validity is subjectivity in prompt tuning. Effective prompts engineering of LLMs are not standardized in software engineering, and hence, prompt configuration is prone to variability. We mitigate this by experimenting with different settings of prompts and iteratively finding the best ones. This again reduces the variability due to prompts, though further refinements are possible.

\textit{\textbf{Conclusion validity}} concerns the validity of the conclusions that is made from the results. To ensure consistency and reduce individual evaluator bias in human evaluation, we used two evaluators for each sample. We also measured interrater reliability using Cohen’s Kappa, which provided an objective metric for agreement among evaluators. Additionally, we provided clear guidelines on evaluation criteria to the evaluators that ensure the reliability of our findings.

\section{Related Work} \label{sec:related-work}
Several approaches have been explored for automatic test case generation, including randomization, search-based algorithms, and deep learning.

Earlier randomization-based approaches, such as Randoop \cite{pacheco2007randoop}, generate test cases based on feedback-directed random sequences of method calls, and adapt based on previous execution results.  Search-based methods like EvoSuite \cite{fraser2011evosuite} employ evolutionary algorithms to iteratively create and optimize test cases for the highest code coverage. Recent works such as AthenaTest \cite{tufano2020unit}, A3Test \cite{alagarsamy2024a3test} frame test case generation as a neural machine translation task (Code→Testcase) and generate test cases for a given (Java) method.

More recently, LLMs have been successfully employed for several software testing tasks, including test case generation \cite{hou2023large, wang2024software}. Schafer et al. designed TestPilot, an adaptive LLM-based test generation tool, to generate unit tests of JavaScript APIs \cite{schafer2023empirical}. Codex \cite{chen2021evaluating} was also utilized to generate code and test cases from natural language descriptions of the Program under Test (PUT) \cite{chen2022codet, lahiri2022interactive}. Chen et al. proposed ChatUniTest, an LLM-based automated unit test generation framework that incorporates an adaptive focal context mechanism to encompass valuable context in prompts and rectify errors in generated unit tests \cite{chen2024chatunitest}. Dakhel et al. proposed Mutation Test case generation with Augmented Prompt (MuTAP) that improves LLM test generation through iterative prompt mutation \cite{dakhel2024effective}. SymPrompt \cite {ryan2024code} structured test suite creation with sequenced, code-aware prompts to improve accuracy and coverage. TestART \cite{gu2024testart} utilized ChatGPT-3.5 based on prompt injection and feedback to improve the reliability of tests and high pass rate, high coverage, thus further demonstrating the potential of LLM-driven testing methods.

Even though LLMs have garnered significant engagement in generating test cases, high-level test case generation using LLMs is a field greatly under-explored, even though such test cases see great usage (Section \ref{sec:high-level-test-case-impact}). Our study addresses this gap by thoroughly exploring the effectiveness of pre-trained LLMs and fine-tuned smaller LLMs in generating high-level test cases. Furthermore, utilizing only the use case of a task to generate test cases through LLMs is also a novel direction. Moreover, unlike most existing techniques (that require code to generate test cases), our proposed dataset and method (leveraging only use cases) supports early testing or test driven development, a key element of many agile frameworks \cite{astels2003test}.

\section{Conclusion and Future Work} 
\label{sec:conclusion}

Our pilot survey of three local software companies points out to one of the most pivotal challenges in software testing: misinterpretation of business requirements and inadequate testing of vital functionalities. To address this, we assessed the potential of high-level test cases: test cases that primarily outline what functionalities and behaviors need to be tested, relaxing focus on the test-case-scripting part. We conducted another broader survey on a diverse collection of software industry experts to understand how helpful or necessary they find these test cases. The survey strongly validated our previous finding that a lack of understanding is the major challenge in software testing. We also discovered that majority of experts rely on high-level test cases and find them useful. They also showed interest in adopting an automated tool for high-level test cases, given that proper confidentiality was ensured.

We rigorously explored the effectiveness of large language models (such as GPT-4o and Gemini) for automatically generating high-level test cases. To address the requirement of confidentiality, we also evaluated fine-tuned smaller open-source models that can be maintained locally. For the evaluation and fine-tuning, we developed a novel dataset containing 3606 test that correspond to 1067 use cases of 168 diffent projects.

To further verify the quality of our curated dataset and the performance of the LLMs, another human survey was performed with the help of software industry experts. The participants evaluated the quality of the test cases generated by humans, pre-trained LLMs, and fine-tuned smaller LLMs on different criteria (e.g, Readability, Correctness, Completeness, Relevance, Usability). While the human-written test cases obtained the highest ratings, scoring very high ($4+$ out of $5$) in each criterion, the LLM-generated ones also received satisfactory scores.

In the future, we plan to-- {i) Analyse LLM performance on challenging project types.} We will identify areas where LLMs underperform, especially in complex or high-stakes applications, to guide model and prompt refinement, {ii) Explore domain-specific fine-tuning} We will fine-tune LLMs using project/domain-specific datasets, such as healthcare, or e-commerce, to improve the generation of relevant test cases for specific industries, {iii) Focus on edge cases.} Curriculum learning  \cite{liu2024let} can be explored in introducing simple cases first and progressing to more complex edge cases to improve the model’s robustness, {iv) Explore requirement generation from project summaries.} We will leverage (and possibly enrich) our dataset to evaluate LLMs in generating detailed requirements and test cases from brief project descriptions, accelerating the initial stages of development.

%Despite our efforts, there is still room for further analysis and enhancement. The LLMs still have limitations in the relevance and completeness criteria, signifying that they sometimes generate irrelevant or repetitive test cases while missing out on some crucial ones. We have partially explored the effectiveness of providing enhanced context in the prompts (Section \ref{sec:enhanced-context}), but much exploring is left to be done through state-of-the-art advanced prompting techniques. Training the LLMs on a larger specialized dataset could also be helpful. Moreover, a multi-stage pipeline of iterative improvement of the high-level test cases could yield more high-quality results.

\section*{Acknowledgement}
This work was funded by the Research and Innovation Center (RIC) under the EDGE Project in Bangladesh.

%\input{acknowledgement}

%\newpage
\begin{small}
\bibliographystyle{abbrv}
%\bibliography{bibtex}
\bibliography{consolidated}
\end{small}

\end{document}